\documentclass[letterpaper,aps,superscriptaddress,showpacs,floatfix,10pt,prc]{revtex4-1}
\usepackage{graphicx}
\usepackage{amsmath}
\usepackage{graphicx}
\usepackage{comment}
\usepackage{float}

\begin{document}
\title{Production of Charge in Heavy Ion Collisions}
\author{Scott Pratt}
\author{William Patrick McCormack}
\affiliation{Department of Physics and Astronomy and National Superconducting Cyclotron Laboratory\\
Michigan State University, East Lansing, MI 48824~~USA}
\author{Claudia Ratti}
\affiliation{Department of Physics, University of Houston, Houston, TX
77204, USA}
\date{\today}


\begin{abstract}
By analyzing preliminary experimental measurements of charge-balance functions from the STAR Collaboration at the Relativistic-Heavy-Ion Collider (RHIC), it is found that pictures where balancing charges are produced in a single surge, and therefore separated by a single length scale, are inconsistent with data. In contrast, a model that assumes two surges, one associated with the formation of a thermalized quark-gluon plasma and a second associated with hadronization, provides a far superior reproduction of the data. A statistical analysis of the model comparison finds that the two-surge model best reproduces the data if the charge production from the first surge is similar to expectations for equilibrated matter taken from lattice gauge theory. The charges created in the first surge appear to separate by approximately one unit of spatial rapidity before emission, while charges from the second wave appear to have separated by approximately a half unit or less.

\end{abstract}

\maketitle

\section{Introduction and Theory}
A principal goal of colliding heavy-ions at high energy is to verify whether one can create small drops of equilibrated quark gluon plasma in the laboratory. Kinetic equilibration is not surprising given the high collision rates and is validated by thermal features of the data, especially observables related to collective flow \cite{Song:2010mg} and jet quenching \cite{Adler:2002tq}. Chemical equilibrium is more difficult to justify. The final-state abundance of hadrons suggests that chemical equilibrium was lost just after hadronization when the temperature was near 165 MeV \cite{BraunMunzinger:2001ip}. In contrast, there is scant experimental evidence that chemical equilibrium was maintained during the period when $T>200$ MeV when the matter is expected to be in a phase of strongly interacting quark gluon plasma (QGP). That chemistry is remarkable. Counting spins, colors and flavors, there are 36 light degrees of freedom from the up, down and strange quarks, and an additional 16 from gluons. Thus, approximately 52 strongly interacting particles should inhabit a volume on the order of one thermal wavelength cubed, $\sim (\hbar c/T)^3$.

At high temperature and zero baryon density, the chemical make-up of the QGP cannot be quantified by counting quarks or gluons because they tend to be off-shell or virtual, so their number is not a well-defined observable. However, even though the average charge within a volume $V$ is zero, the fluctuations of the charge characterize the degrees of freedom. If one considers the three-by-three fluctuation tensor,
\begin{equation}\label{eq:chidef}
\chi_{ab}=\frac{Q_aQ_b}{V},
\end{equation}
with $a$ and $b$ referring to the up, down and strange charge, one can gain insight into the chemistry. If up, down and strange quarks are good quasi-particles with no inter-quark correlations (a gaseous state),
\begin{eqnarray}\label{eq:chiqgp}
\chi_{ab}^{\rm(QGP)}=(n_a+n_{\bar{a}})\delta_{ab},
\end{eqnarray}
where $n_a$ is the density of quarks of the given flavor, up, down or strange. Correlations between quarks, such as the two being in the same hadron, alter the expression. For a hadron gas the correlations are
\begin{eqnarray}\label{eq:chihad}
\chi_{ab}^{\rm(had)}=\sum_\alpha n_{\alpha}q_{\alpha a}q_{\alpha b},
\end{eqnarray}
where $n_\alpha$ is the density of hadron species $\alpha$ which has a charge $q_{\alpha a}$. As an example, protons contribute a factor of four times their density to $\chi_{uu}$ because the ``up'' charge of a proton is two. Even for strongly interacting and correlated systems $\chi_{ab}$ is a well-defined observable, because charge is conserved by the strong interaction. From equation \ref{eq:chihad} one can see that hadronic resonances induce off-diagonal elements to $\chi_{ab}$ \cite{Koch:2005vg}. For example, the $K^+$ contributes negatively to $\chi_{us}$ and a $\Lambda$ hyperon gives a positive contribution to $\chi_{us}$. 

\begin{figure}
\centerline{\includegraphics[width=0.35\textwidth]{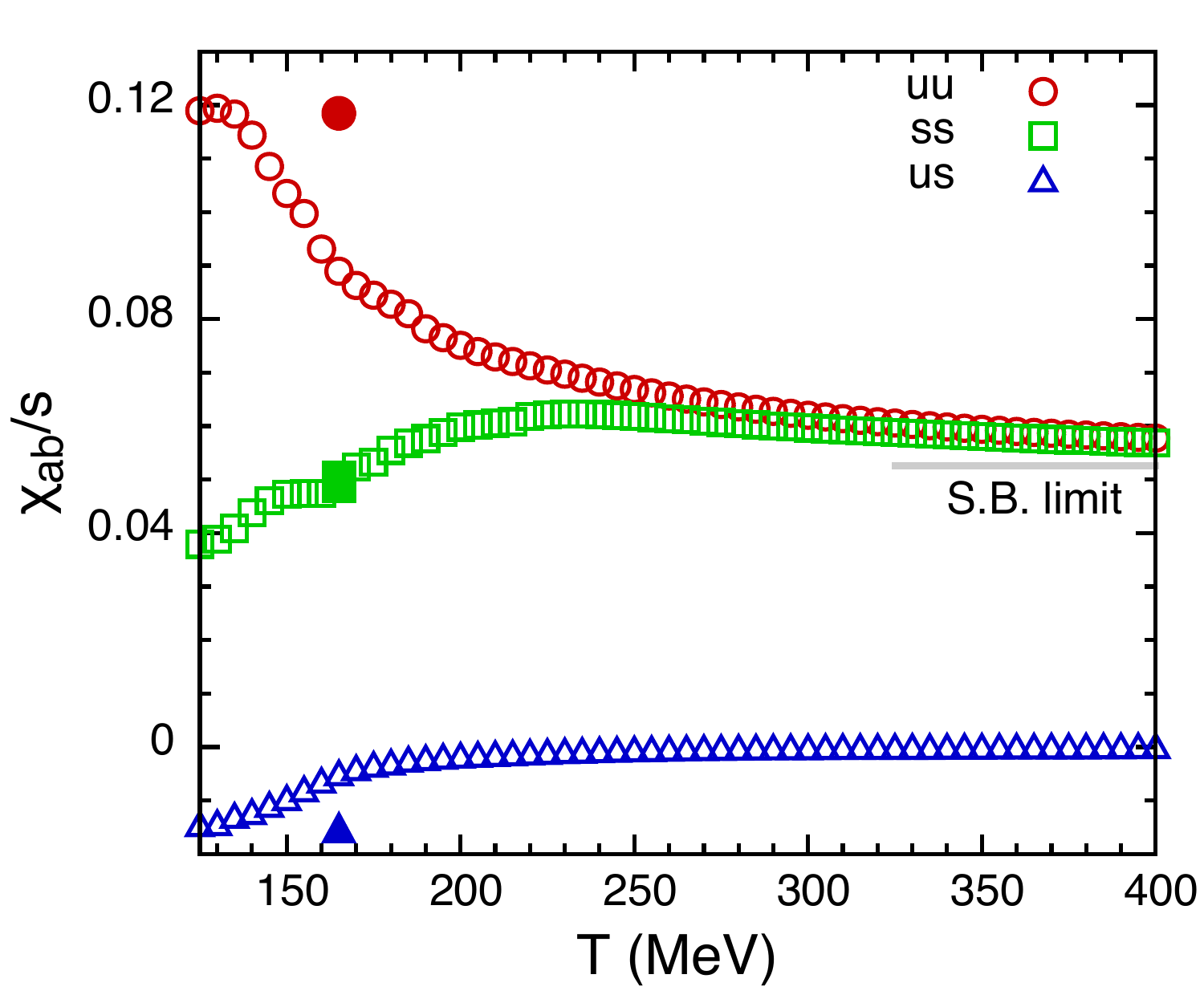}}
\caption{\label{fig:lattice}
Charge fluctuations from lattice gauge theory \cite{Borsanyi:2011sw,Ratti:2011au} (open symbols) are similar to those of a hadronic gas (filled symbols for $T=165$ MeV). For fixed entropy there are increased numbers of up and down quarks in the hadronic phase, whereas the number of strange quarks is slightly smaller. The off-diagonal element disappears above $T_c$ when hadrons dissolve and quark-antiquark correlations disappear. At high temperatures the results approach those of a Stefan-Boltzmann gas of massless partons (S.B. limit).
}
\end{figure}

For a massless gas of quarks and gluons, the number of quarks within a fluid element of fixed entropy stays constant during an isentropic expansion because both the number densities and the entropy density scale as $T^3$. Therefore, in an isentropic expansion of a massless parton gas the number of quarks within the fluid element stays relatively constant. Lattice calculations show that this property remains reasonably preserved even for a strongly interacting system. The ratio $\chi_{ab}/s$ is illustrated in Fig. \ref{fig:lattice} where it changes only at the 10\% level once $T> 225$ MeV. Below that temperature, the system is hadronizing and $\chi$ is strongly temperature dependent. Due to entropy conservation, the number of hadrons just below $T_c$ is marginally lower than the number of quarks above $T_c$, so given that each hadron has two or three quarks, copious quark production, through string breaking or resonance decays, accompanies hadronization. As shown in Fig. \ref{fig:lattice}, some of the elements of the fluctuation tensor change significantly, especially $\chi_{uu}/s\approx\chi_{dd}/s$ which nearly doubles. 

Unfortunately, charge fluctuations, $\chi_{ab}(T)$, are not directly accessible from experiment. If one could measure the charge fluctuation within a small volume, a few fm$^3$, one would expect the fluctuation to approximate that of an equilibrated system as long as the equilibrated system does not have correlations beyond that scale. However, if one considers the net charge in the entire system, it does not fluctuate due to the fact that, unlike the assumptions for a grand canonical ensemble, the net charge is fixed and $\chi\rightarrow 0$. In contrast, the charge correlation 
\begin{equation}
g_{ab}(\Delta \eta)\equiv\left\langle\rho_a(0)\rho_b(\Delta\eta)\right\rangle,
\end{equation}
is sensitive to both the equilibrated charge fluctuation and accounts for the conservation of net charge. Here, $\rho_a(\eta)$ is the charge density per unit of spatial rapidity and the brackets denote averages over events. Translational invariance along the beam axis, or invariance to translations in spatial rapidity, is assumed. Spatial rapidity is a measure of the position along the beam, or $z$, axis, $z/t=\tanh \eta$. If all fluid elements begin at $z=t=0$ and if the elements do not accelerate longitudinally, which is expected for a boost-invariant system, the spatial rapidity can be associated with the longitudinal velocity of the fluid element, $v_z=z/t$. If a particle moves with the fluid, its spatial rapidity stays constant \cite{Bjorken:1982qr}.

If chemical equilibrium is attained and if the correlations are local the correlation becomes
\begin{equation}
g_{ab}(\Delta\eta)\rightarrow\chi_{ab}\delta(\Delta\eta),
\end{equation}
where the $\delta$ function can be relaxed to some function of short range that integrates to unity. For the realistic situation where the charge is created locally and diffuses over a finite distance, the correlation becomes
\begin{eqnarray}
\label{eq:chargeconservation}
g_{ab}(\Delta\eta)&=&\chi_{ab}\delta(\Delta\eta)+g'_{ab}(\Delta\eta),\\
\nonumber
\int d\Delta\eta~ g'_{ab}(\Delta\eta)&=&-\chi_{ab}.
\end{eqnarray}
The last condition derives from charge conservation.

Even though the net charge over the entire collision volume does not fluctuate and even though the short-range correlation only carries information about the current value of $\chi$, one can gain insight into the temporal history of $\chi_{ab}$ by analyzing the dependence of $\chi_{ab}$ on $\Delta\eta$, or in a three dimensional analysis on $\Delta x,\Delta y,\Delta \eta$. For example, if $\chi_{ab}/s$ were to stay constant after the initial time, $g'_{ab}(\Delta\eta)$ would be a broad function with a width determined by how far charge pairs produced early would separate over the history of the collision. This width might be driven by a combination of two effects. First, if the balancing quarks are produced by the fragmentation of a longitudinal flux tube, the tunneling would lead to the charges being significantly separated at birth along the beam direction, perhaps by a solid fraction of one unit of spatial rapidity. For example, if the quarks were born with a separation of 0.5 fm, and  were created at a proper time of 0.5 fm/$c$, the separation would be on unit of spatial rapidity. In contrast, if particles were born 0.5 fm apart at a time of 5 fm/$c$, the separation would be only 0.1 units of spatial rapidity. If the system maintains local chemical equilibrium, as defined by the  lattice calculations displayed in Fig. \ref{fig:lattice}, at hadronization a second contribution to $g_{ab}$ would arise to account for the change in $\chi_{ab}$. This second contribution to $g'_{ab}(\Delta\eta)$ would be more tightly constrained in $\Delta\eta$ due to the reduced time available for the charges to diffuse away from one another. 

Diffusion represents another mechanism of separation. For a strongly interacting QGP the diffusion constant and the separation would tend to be smaller. From \cite{Bass:2000az} one can estimate the diffusive separation with a simple analytic formula based on assuming cross sections stayed constant with temperature. The diffusive width to the balance function was
\begin{equation}
\sigma_\eta^2=4\beta\ln(\tau/\tau_0), ~~~\beta= v_t/(n\tau\sigma).
\end{equation}
If the cross section of $\sigma=10$ mb = 1.0 fm$^2$ were used, and if the density scaled by the time $\tau$ were $n\tau=5$ fm$^{-2}$, and if $\tau/\tau_0=10$, the particles would separate by approximately one unit of spatial rapidity. If several collisions were required to re-thermalize a particle the diffusive separation would increase. 

The correlations in terms of charges, $g_{ab}$, translate into correlations between specific hadronic species $\alpha$ and $\beta$,,
\begin{equation}
G_{\alpha\beta}(\Delta\eta)\equiv \langle (n_{\alpha}(0)-n_{\bar{\alpha}}(0))
(n_{\beta}(\Delta\eta)-n_{\bar{\beta}}(\Delta\eta))\rangle.
\end{equation}
A relationship between $g_{ab}(\Delta\eta)$ and $G_{\alpha\beta}(\Delta\eta)$ can be derived if one assumes that a small balancing charge is spread randomly, or thermally, amongst the hadrons \cite{Pratt:2012dz}.

Using a blastwave prescription, which provides a parametric description of thermal emission overlaid with collective flow and includes the effects of decays, one can then project the correlations in coordinate space to correlations in momentum space. These correlations are divided by the yield of particles of type $\alpha$ and are called charge balance functions,
\begin{equation}
B_{\alpha\beta}(\Delta y)=\frac{ \langle (n_{\alpha}(0)-n_{\bar{\alpha}}(y))
(n_{\beta}(0)-n_{\bar{\beta}}(\Delta\eta))\rangle}{\bar{n}_\alpha+\bar{n}_{\bar{\alpha}}},
\end{equation}
which have been measured by experiment \cite{Adamczyk:2015yga,Wang:2011za,STAR:2011ab,Aggarwal:2010ya,Abelev:2010ab,Westfall:2004cq,Westfall:2004jh,Adams:2003kg,Weber:2013fla,Abelev:2013csa,Weber:2012ut}.

In the next section, the methods are reviewed for translating charge correlations, $g_{ab}(\Delta\eta)$, to $G_{\alpha\beta}(\Delta\eta)$ and then to balance functions $B_{\alpha\beta}(\Delta y)$ using a blast-wave prescription. Preliminary measurements from STAR are reviewed in Sec. \ref{sec:star}, while Sec.s \ref{sec:datavsmodel} and \ref{sec:stats} illustrate how individual balance functions are determined by specific parameters in the model. In Sec. \ref{sec:stats} eight model parameters are systematically varied in a Markov Chain Monte Carlo (MCMC) exploration of parameter space that implements model emulator techniques. The results make a strong case that the matter created in central Au+Au collisions at RHIC is close to being chemically equilibrated early in the collision. In Sec. \ref{sec:summary} these findings are summarized and some ideas for future analysis are presented.

\section{Generalized Balance Functions and the Two-Surge Blast-Wave Model}
\label{sec:twosurgemodel}

In this paper we present comparisons to a simple model that can reproduce the behavior expected from the trends seen in the lattice results displayed in Fig. \ref{fig:lattice}. The picture assumes two surges of charge production: one where the initial equilibrated matter would be formed and a second corresponding to the jump in the susceptibilities near $T_c$. To that end, the model will assume that the first surge results in charge correlations significantly spread in rapidity, while the second surge, occurring as the temperature falls below $T\sim 200$ MeV, would result in significantly more compact correlations. The spread in spatial rapidity at breakup from the two surges will be described by two widths, $\sigma_A$ for the first surge and $\sigma_B$ for the second surge. First, we repeat the derivations in \cite{Pratt:2012dz} and describe how these correlations carry over into correlations in rapidity between two specific hadronic species.

The $3\times 3$ charge correlation matrix in coordinate space, $g'_{ab}(\Delta\eta)$, determines the observable hadronic correlations in the final state if one assumes that the differential additional charges in a small volume are distributed thermally amongst the hadrons. Here, we define a correlation between two hadronic species $\alpha$ and $\beta$ \cite{Pratt:2012dz},
\begin{eqnarray}
\label{eq:Gdef}
G_{\alpha\beta}(\Delta\eta)&\equiv& \langle (n_{\alpha}(0)-n_{\bar{\alpha}}(0))
(n_{\beta}(\Delta\eta)-n_{\bar{\beta}}(\Delta\eta))\rangle,\\
g_{ab}(\Delta\eta)&=&\frac{1}{4}\sum_{\alpha\beta}G_{\alpha\beta}(\Delta\eta)q_{\alpha a}q_{\beta b},
\end{eqnarray}
where $q_{\alpha a}$ is the charge of type $a$ carried by the hadron species $\alpha$. There are many more hadronic species than charges, so $G_{\alpha\beta}$ cannot be uniquely determined from $g_{ab}$ without making assumptions. The factor of $1/4$ accounts for the double counting of species. For instance, the sum over $\alpha,\beta$ includes both $\pi^+\pi^-$ and $\pi^-\pi^+$.

If one assumes that an additional differential charge density $\delta \rho_a$ within a given volume element is spread amongst the various species randomly, i.e. thermally, the change in each species' yield can be found by finding the chemical potential $\delta\mu_a$ required to generate $\delta \rho_a$. 
\begin{eqnarray}
\delta n_\alpha&=&(e^{\delta\mu_a q_{\alpha a}}-1)\bar{n}_\alpha\\
\nonumber
&=&\delta\mu_a q_{\alpha a}\bar{n}_{\alpha},
\end{eqnarray}
where $\delta\mu$ has absorbed the $1/T$ factor typically used in defining chemical potentials and is dimensionless, while $\bar{n}_\alpha$ refers to the average particle density. The three values $\delta\mu_a$ can now be found from the three constraints in Eq. (\ref{eq:Gdef}). In turn these then determine $\delta n_{\alpha}$. 
\begin{eqnarray}
\nonumber
\delta\rho_a&=&\sum_\alpha q_{\alpha a} \sum_b \bar{n}_{\alpha}q_{\alpha b}\delta\mu_b ,\\
\nonumber
&=&\sum_b\chi^{\rm(had)}_{ab}\delta\mu_b,\\
\nonumber
\delta\mu_a&=&\sum_b\chi^{{\rm(had)}-1}_{ab}\delta\rho_b,\\
\delta n_\alpha&=& \bar{n}_\alpha \sum_{ab}q_{\alpha a} \chi^{{\rm(had)}-1}_{ab}\delta\rho_b.
\end{eqnarray}
Using the fact that in Eq. (\ref{eq:Gdef}) $n_{\alpha}-n_{\bar{\alpha}}=2\delta n_\alpha$, 
\begin{equation}
\label{eq:Gexpression}
G_{\alpha\beta}(\Delta\eta)=4\sum_{abcd}
\bar{n}_\alpha q_{\alpha a} \chi^{{\rm(had)}-1}_{ab} g_{bc}(\Delta\eta) \chi^{{\rm(had)}-1}_{cd} q_{\beta d}\bar{n}_\beta.
\end{equation}
Using the expression for  $\chi^{\rm(had)}$ in Eq. (\ref{eq:chihad}) one can see that the form is consistent,
\begin{equation}
\frac{1}{4}\sum_{\alpha\beta}G_{\alpha\beta}(\Delta\eta) q_{\alpha a}q_{\beta b} = g_{ab}(\Delta\eta).
\end{equation}

In the two-surge model, calculating the experimentally measurable $G_{\alpha\beta}(\Delta y)$, where $\Delta y$ is the correlation in relative asymptotic rapidity, involves the following steps:
\begin{enumerate}\itemsep=0pt
\item {\bf Calculate $g_{ab}(\Delta\eta)$ at hadronization}. Assume a form consisting of two surges, 
\begin{eqnarray}
\label{eq:g2wave}
g'_{ab}(\Delta\eta)&=&
-\left(\chi^{\rm(had)}_{ab}-\chi^{\rm(QGP)}_{ab}\right)\frac{e^{-\Delta\eta^2/2\sigma_B^2}}{(2\pi\sigma_B^2)^{1/2}}\\
\nonumber
&&-\chi^{\rm(QGP)}_{ab}\frac{e^{-\Delta\eta^2/2\sigma_A^2}}{(2\pi\sigma_A^2)^{1/2}}
\end{eqnarray}
Here, $\sigma_A$ describes the width in spatial rapidity to which charges created in the first wave have separated by the time one reaches breakup and $\sigma_B$ characterizes the separation from the second wave at breakup. The strength of the first surge, $\chi_{ab}^{\rm(QGP)}$, is the susceptibility using a number per unit rapidity, or taken from lattice,
\begin{equation}
\chi_{ab}^{\rm(QGP)}=\frac{\chi_{ab}^{\rm(lattice)}(T\approx 300~{\rm MeV})}{s} \frac{dS}{d\eta},
\end{equation}
and $\chi_{ab}^{\rm(had)}$ is given by the properties at chemical freezeout, which for this calculation will be defined by a temperature $T_{\rm chem}=165$ MeV. One can also calculate the entropy density of a hadron gas at chemical equilibrium, $s_{\rm chem}$. 
\item $G_{\alpha\beta}(\Delta\eta)$ is calculated from $g_{ab}$ using Eq. (\ref{eq:Gexpression}). After substituting Eq. (\ref{eq:g2wave}) into Eq. (\ref{eq:Gexpression}),
\begin{eqnarray}
\label{eq:Gtwoscale}
G_{\alpha\beta}(\Delta\eta)&=&-w_{A,\alpha\beta}
\bar{n}_\alpha\bar{n}_\beta\frac{e^{-\Delta\eta^2/2\sigma_A^2}}{(2\pi\sigma_A^2)^{1/2}}
-w_{B,\alpha\beta}\bar{n}_\alpha\bar{n}_\beta
\frac{e^{-\Delta\eta^2/2\sigma_B^2}}{(2\pi\sigma_B^2)^{1/2}},\\
\nonumber
w_{A,\alpha\beta}&=&4\sum_{ab}
q_{\alpha a}\left(\chi^{{\rm(had)}-1}\chi^{\rm(QGP)}\chi^{{\rm(had)}-1}\right)_{ab} q_{\beta b},\\
\nonumber
w_{B,\alpha\beta}&=&4\sum_{ab}q_{\alpha a}\chi^{{\rm(had)}-1}_{ab}q_{\beta b}-w_{A,\alpha\beta}.
\end{eqnarray}
\item Given $G_{\alpha\beta}(\Delta\eta)$ at freezeout, the correlations in coordinate space are then mapped to correlations in spatial rapidity, $G_{\alpha\beta}(\Delta y)$, according to the blast-wave prescription. This mapping is done by the following Monte Carlo procedure. (a) Pairs of particles are chosen with probability proportional to $\bar{n}_\alpha\bar{n}_\beta$. (b) The particles are then placed in coordinate space with the relative position randomly according to a Gaussian width $\sigma_A$. (c) The particle's local momenta are generated thermally, then boosted transversely according to the blast-wave prescription described below.  (d) If the particles are both stable the balance functions for the given relative rapidity bin is incremented by $w_{A,\alpha\beta}\bar{n}_{\rm tot}^2\epsilon_\alpha\epsilon_\beta$, where $\bar{n}_{\rm tot}$ is the number of hadrons per unity rapidity and $\epsilon_\alpha$ is the efficiency with which particle $\alpha$ is measured, and depends on the particle type, its transverse momentum and rapidity\cite{staracceptance} (e) If one or both particles are unstable, they are decayed and the relative rapidity is calculated for each of the products $\alpha'$ and $\beta'$. Each correlation $G_{\alpha'\beta'}$ is then incremented by $w_{A,\alpha\beta}\bar{n}_{\rm tot}^2\epsilon_{\alpha'}\epsilon_{\beta'}$. Steps b-f are repeated using the weight $w_B$ and the width $\sigma_B$. Decays are accounted for by considering the intra-correlations between any of the particles of type $A$ and species $\alpha'$ and $\alpha''$ and using a weight $\bar{n}_\alpha$.
\item After using the previous steps to calculate $G_{\alpha\beta}(\Delta y)$, one can then calculate the ``generalized'' balance function, i.e. the balance function for specific species, by the relation
\begin{equation}
B_{\alpha\beta}(\Delta y)=\frac{G_{\alpha\beta}(\Delta y)}{\bar{n}_\alpha+\bar{n}_{\bar{\alpha}}}.
\end{equation}
The denominator requires generating particles with the species being picked proportional to $\bar{n}_\alpha$. Then one should boost and decay the particles, and increment the denominator for the type $\alpha'$ by the efficiency for the final species by an amount $\epsilon_{\alpha'}n_{\rm(tot)}$. As long as the number of samplings is the same as the previous step, the normalization should be correct.

One can test calculations along the way by considering the normalization of various quantities with perfect acceptance. 
\begin{eqnarray}
\sum_\beta (w_{A,\alpha\beta}+w_{B,\alpha\beta})\bar{n}_\beta q_{\beta a}&=&4q_{\alpha a},\\
\nonumber
\sum_{\alpha\beta}\int d\Delta\eta~q_{\alpha a}G_{\alpha\beta}(\Delta\eta)q_{\beta b}&=&-4\chi_{ab},\\
\nonumber
\sum_\beta\int d\Delta y~B_{\alpha\beta}(\delta y)q_{\beta b}&=&-2q_{\alpha b},\\
\end{eqnarray}
The quantities also have the symmetry, $G_{\alpha\bar{\beta}}=-G_{\alpha\beta}$ and $B_{\alpha\bar{\beta}}=-B_{\alpha\beta}$.
\end{enumerate}

The blast wave used to transversely boost the particles uses four parameters. The first is the kinetic temperature at breakup. Although particle yields are determined by the temperature $T_{\rm chem}=165$ MeV, where chemical equilibrium would be lost, the particle's momenta are determined by the kinetic temperature $T_{\rm kin}=102$ MeV. The second parameter is the transverse collective velocity, $u_\perp=0.732$. After being generated thermally, they are boosted by a transverse velocity chosen from the distribution, 
\begin{equation}
\label{eq:uperp}
\frac{dN}{du_xdu_y}\sim e^{-(u_x^2+u_y^2)/2u_\perp^2}.
\end{equation}
The two parameters $u_\perp$ and $T_{\rm kin}$ are chosen to roughly reproduce the mean transverse momentum of pions and protons seen by the STAR \cite{Abelev:2008ab} and PHENIX \cite{Adler:2003cb} collaborations at RHIC. The third and fourth parameters determine the chemistry. In addition to $T_{\rm chem}$, a parameter $F_B=2/3$ is used to reduce the yield of all baryons to account for baryon annihilation below $T_{\rm chem}$ \cite{Pan:2012ne,Steinheimer:2012rd}. Rather than employing the rather large baryon annihilation factors here, calculations were also performed using lower values of $T_{\rm chem}$  as suggested in \cite{Ratti:2011au,Borsanyi:2014ewa,Bluhm:2014wha}. If the final $p/\pi$ ratios are similar, results change little. 

The blast wave picture used here is perhaps the simplest model that can readily incorporate two surges while providing a reasonable mapping between the correlations in coordinate space and those in asymptotic relative rapidity. If both widths $\sigma_A$ and $\sigma_B$ are set equal to one another, it becomes a one-surge model. Of course, both charge creation and emission are more complicated than what can be described in this picture. First, although one expects two surges, the susceptibility seen in Fig. \ref{fig:lattice} does rise slowly as the temperature falls in the region $T\gtrsim 200$ MeV. One could replace the two surges with a continuum of creation proportional to $d/d\tau(\chi_{ab}/s)$. Each point in space would then contribute to the final $g_{ab}$. However it is clear that, if the system stays close to chemical equilibrium, the bulk of these contributions comes from two surges. Additionally, if either surge is not strongly confined to a specific time, the shape of the ensuing correlation should no longer be Gaussian, even if the separation is purely diffusive. A second weakness of the model comes from the assumption that the matter disassociates simultaneously according to a single breakup temperature, $T_{\rm kin}$, but with yields determined by a single temperature $T_{\rm chem}$. For shorter lived resonances like the $\rho$ meson, yields should substantially drop between the two temperatures. Thus, the contributions from resonance decays are probably significantly overstated. However, if a neutral particle decays and rethermalizes, the ensuing relative rapidity of the two charges is not wholly different than if the two decay prodcucts escaped unscathed, because these decays have relative momenta of typical thermal scales. Despite these shortcomings, the two-surge model should provide an insightful point of comparison. A more realistic picture of creation, transport and emission of conserved charges is currently being pursued by some of the authors.

\section{Summarizing Experimental Measurements}
\label{sec:star}

The STAR Collaboration has presented preliminary balance functions for four combinations of species: $\pi^+\pi^-$, $K^+K^-$, $p\bar{p}$ and $K^-p$ \cite{Wang:2012jua,WestfallWPCF}, displayed in Fig. \ref{fig:star}. Before discussing the result, we apply an approximate acceptance and efficiency correction so that the true widths of the balance functions can be better discerned. This correction is approximate and based on an assumption that the balance functions depend only on the relative rapidity, or equivalently that the balance functions do not depend on the transverse momenta of the two particles. Assuming equal numbers of particles and antiparticles, the balance function can be exactly expressed as
\begin{eqnarray}
\label{eq:Bacc1}
B(\Delta y)&=&\frac{\int dp_ady_adp_bdy_b~\delta(|y_b-y_a|-\Delta y)
\left[N_{\alpha\beta}(p_a,y_a,p_b,y_b)-N_{\alpha\bar{\beta}}(p_a,y_a,p_b,y_b)\right]}
{\int dp_ady_a~N_\alpha(p_a,y_a)},
\end{eqnarray}
with $p_a$ referring to the transverse momentum coordinates of the particle of type $\alpha$ and $p_b$ doing the same for type $\beta$. The rapidities of the two types are then $y_a$ and $y_b$. In terms of the balance function,
\begin{eqnarray}
\label{eq:Bacc2}
N_{\alpha\beta}(p_a,y_a,p_b,y_b)-N_{\alpha\bar{\beta}}(p_a,y_a,p_b,y_b)&=&
N_\alpha(p_a,y_a)B^{\rm(p)}(p_a,y_a,p_b,y_b)A_\beta(p_b,y_b),
\end{eqnarray}
where $A_\beta(p_b,y_b)$ is the efficiency that can vary between zero and unity, and $B^{\rm(p)}$ is the balance function if the acceptance for the particle of type $\beta$ were perfect. Now, to make the approximation, one assumes that an additional particle of type $\beta$ at $y_b$ has its transverse momentum distribution distributed according to the single-particle probability, $N_\beta(p_b,y_b)$, with half the strength coming at $y_b=y_a+\Delta y$ and the other at $y_b=y_a-\Delta y$. 
\begin{eqnarray}
\label{eq:Bacc3}
B^{\rm(p)}(p_a,y_a,p_b,y_b)&\approx&\frac{1}{2}
B^{\rm(p)}(\Delta y)\frac{N^{\rm(p)}_\beta(p_b,y_b)}{\int dp_b~N^{\rm(p)}_\beta(p_b,y_b)}.
\end{eqnarray}
One can now insert Eq.s (\ref{eq:Bacc2}) and (\ref{eq:Bacc3}) into Eq. (\ref{eq:Bacc1}) and see that $B^{\rm(p)}(\Delta y)$ factors out of the expression,
\begin{eqnarray}
B(\Delta y)&=&B^{\rm(p)}(\Delta y)
\left\{\frac{\int dp_ady_a N_\alpha(p_a,y_a)\bar{A}(y_b=y_a+\Delta y)}
{2\int dp_ady_a~N_\alpha(p_a,y_a)}\right.\\
\nonumber
&&\hspace*{70pt}\left.+\frac{\int dp_ady_a N_\alpha(p_a,y_a)\bar{A}(y_b=y_a-\Delta y)}
{2\int dp_ady_a~N_\alpha(p_a,y_a)}\right\}~,\\
\nonumber
\bar{A}_\beta(y_b)&\equiv&\frac{\int dp_b~N_\beta^{\rm(p)}(p_b,y_b)A_\beta(p_b,y_b)}
{\int dp_b~N_\beta^{\rm(p)}(p_b,y_b)}
\end{eqnarray}
If the efficiency $A_\beta(p_b,y_b)$ is perfect, the balance function is $B^{\rm(p)}(\Delta y)$. 
The expression in the brackets on the r.h.s. represents the acceptance correction, $C(\Delta y)$.
\begin{eqnarray}
B^{\rm(p)}(\Delta y)&=&\frac{B(\Delta y)}{C(\Delta y)}~,\\
\nonumber
C(\Delta y)&=&\frac{\int dp_ady_a N_\alpha(p_a,y_a)\bar{A}_\beta(y_a+\Delta y)}
{2\int dp_ady_a~N_\alpha(p_a,y_a)}
+\frac{\int dp_ady_a N_\alpha(p_a,y_a)\bar{A}_\beta(y_a-\Delta y)}
{2\int dp_ady_a~N_\alpha(p_a,y_a)}~.
\end{eqnarray}
The acceptance probability $A_\beta(y_b)$ is basically the ratio of the measured yield of particles of type $\beta$ at $y_b$ relative to the true yield. Using a STAR acceptance and efficiency filter \cite{staracceptance}, this can be found by generating particles with the blast-wave prescription, then seeing what fraction are recorded. The correction factor $C(\Delta y)$ then requires averaging $A_\beta$ over the various possibilities for $y_a$, which can again be performed with the blast-wave prescription. Figure \ref{fig:star} shows both the original preliminary balance function from STAR and the corrected version. For large $\Delta y$ the factor is large and the experimental uncertainties are magnified. Points are not plotted where the uncertainty surpasses the maximum size of the balance function.

\begin{minipage}{0.475\textwidth}
\begin{figure}[H]
\centerline{\includegraphics[width=0.7\textwidth]{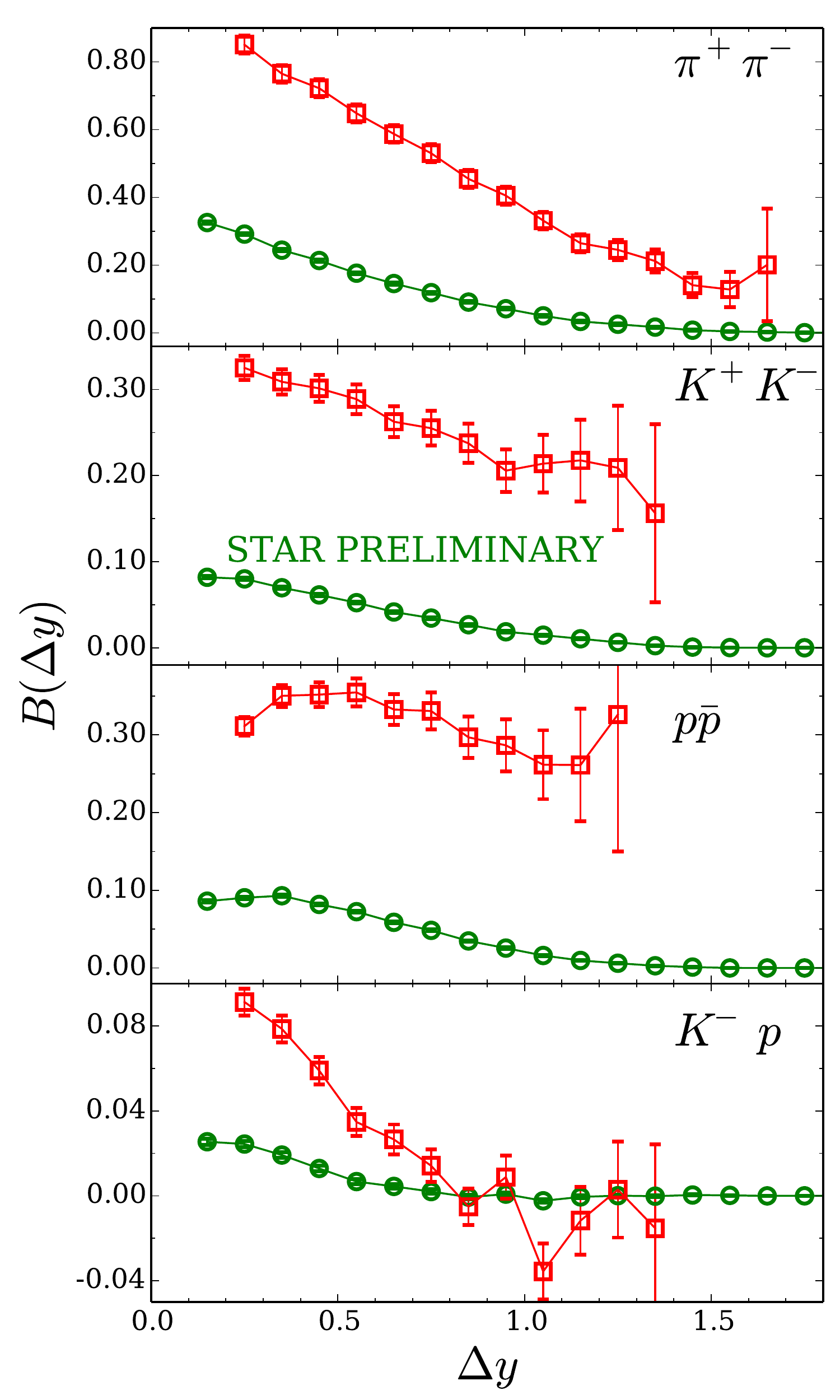}}
\caption{\label{fig:star}
Preliminary balance functions from STAR (green circles) are shown from four pairs of species: $\pi^+\pi^-$, $K^+K^-$, $p\bar{p}$ and $K^-p$. The upper curves (red squares) show the result after applying the approximate acceptance and efficiency correction described here. Measurements are for the most central collisions, 0-5\%.}
\end{figure}
\end{minipage}
\hspace*{0.05\textwidth}
\begin{minipage}{0.475\textwidth}
\begin{figure}[H]
\centerline{\includegraphics[width=0.7\textwidth]{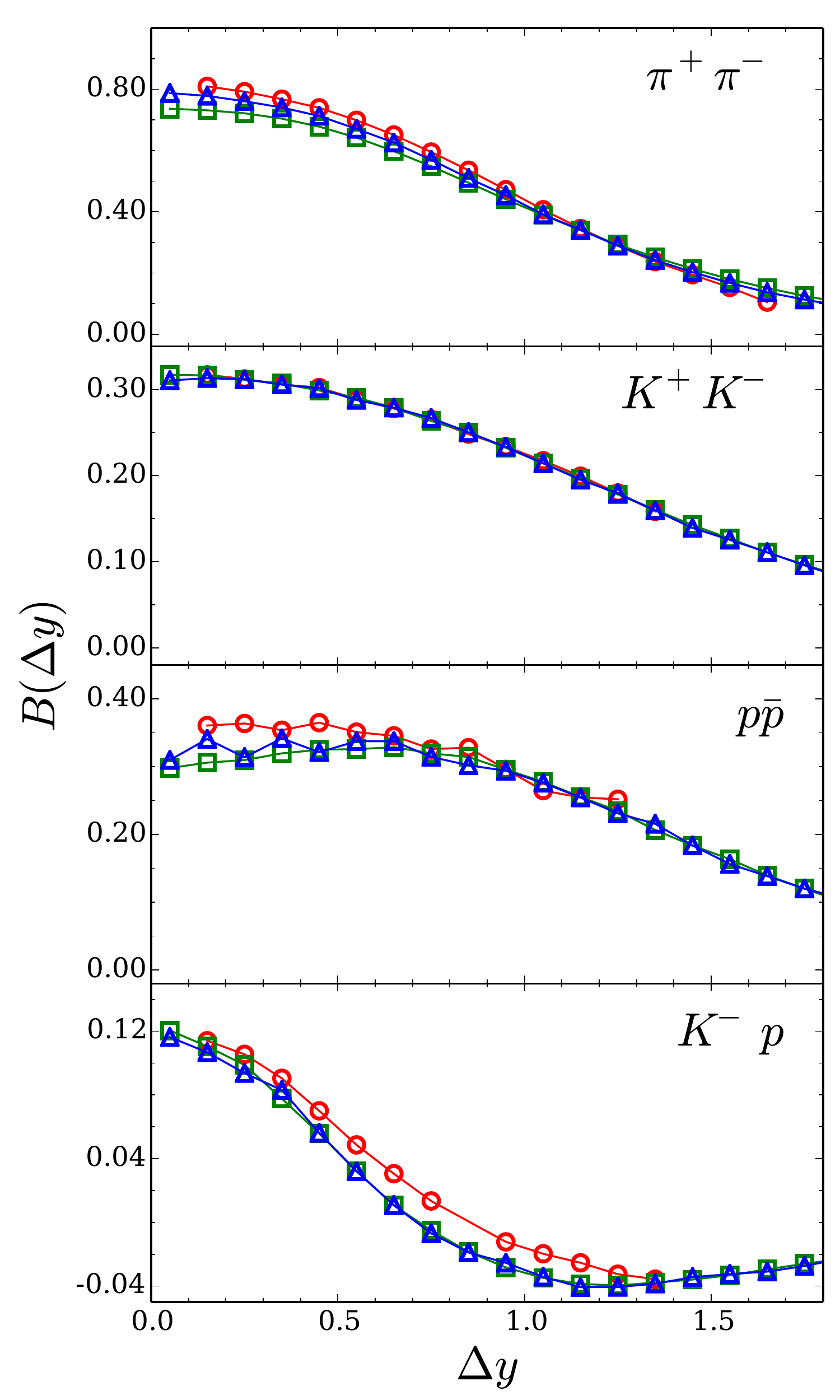}}
\caption{\label{fig:acceptance}
The acceptance and efficiency corrected balance funtion, $B_{\alpha\beta}(\Delta y)$, from model calculations (red circles) is compared to the corresponding calculations with the acceptance and efficiency for particle type $\beta$ being perfect (blue triangles) and for both particles having perfect acceptance (green squares). If the assumptions, upon which the acceptance and efficiency corrections were based, are justified, the three calculations would be nearly identical.
}
\end{figure}
\end{minipage}
\medskip

Because the form for the acceptance and efficiency corrections was built on what may be a dubious assumption, they were also applied to model balance functions. In this case, the corrected balance functions can be compared to those calculated with perfect acceptance to see whether the difference between corrected and perfect balance functions differ. Figure \ref{fig:acceptance} shows three sets of balance functions $B_{\alpha\beta}(\Delta y)$, the corrected version as described above, the balance function where the efficiency for the particle of type $\beta$ is perfect but where the acceptance and efficiency for $\alpha$ is given by the filter, and finally a balance function where the acceptance for both types is perfect. Because a balance function's denominator divides out the efficiency of the first particle, all three would be identical if the balance function were truly only a function of $\Delta y$ as assumed in the approximation described by Eq. (\ref{eq:Bacc3}). The parameters used in the model were  for the two-surge model, chosen to roughly reproduce the experimental balance functions: $\sigma_A=1.0$, $\sigma_B=1/3$. The susceptibility for the quark gluon plasma was chosen to be $\sum_a\chi_{ab}/s=0.18$ and $\chi_{ss}/\chi_{uu}=0.93$, to be roughly consistent with the lattice calculations in Fig. \ref{fig:lattice} for temperatures near 300 MeV. 

Even without a model the STAR measurements make it clear that describing the separation of balancing charges with a single scale in relative spatial rapidity cannot describe the data. The four experimental balance functions for the most central collisions are displayed in Fig. \ref{fig:star}. In the absence of acceptance corrections, a one-surge model, such as Eq. (\ref{eq:Gtwoscale}) with $\sigma_A=\sigma_B$, would give balance functions of equal width if they were calculated as a function of spatial rapidity. However, given that the mapping to spatial rapidity is thermally smeared to a larger extent for lighter particles due to their higher thermal velocity, one would expect the following hierarchy of widths, $\sigma_{\alpha\beta}$: 
\begin{equation}
\sigma_{\pi^+\pi^-}>\sigma_{K^+K^-}>\sigma_{pK^-}>\sigma_{p\bar{p}}.
\end{equation}
The STAR data violates this expectation as it appears that
\begin{equation}
\sigma_{p\bar{p}}>\sigma_{K^+K^-}>\sigma_{\pi^+\pi^-}>\sigma_{pK^-}.
\end{equation}
The $\pi^+\pi^-$ balance function should be strongly affected by decays. If one were to decay a chemically equilibrated hadron gas at $T=165$ without any further hadronic interactions, nearly half the positive pions would come from decays where the $\pi^+$ is accompanied by a $\pi^-$. Thus, it should not be surprising that the $\pi^+\pi^-$ balance function has a relatively narrow component. This could either be considered a third surge, or as an extension of the second surge. Determining what part of this narrowing is due to decays vs. the rise in the charge susceptibility in the transition region requires detailed analysis. However, the fact that the $K^-p$ balance function is narrower than either the $p\bar{p}$ or $K^+K^-$ balance functions, and even narrower than the $\pi^+\pi^-$ balance functions, seems impossible to describe with a one-surge model.

\section{Comparing the Two-Surge Model to Preliminary STAR Results}
\label{sec:datavsmodel}

\begin{figure}
\centerline{\includegraphics[width=0.6\textwidth]{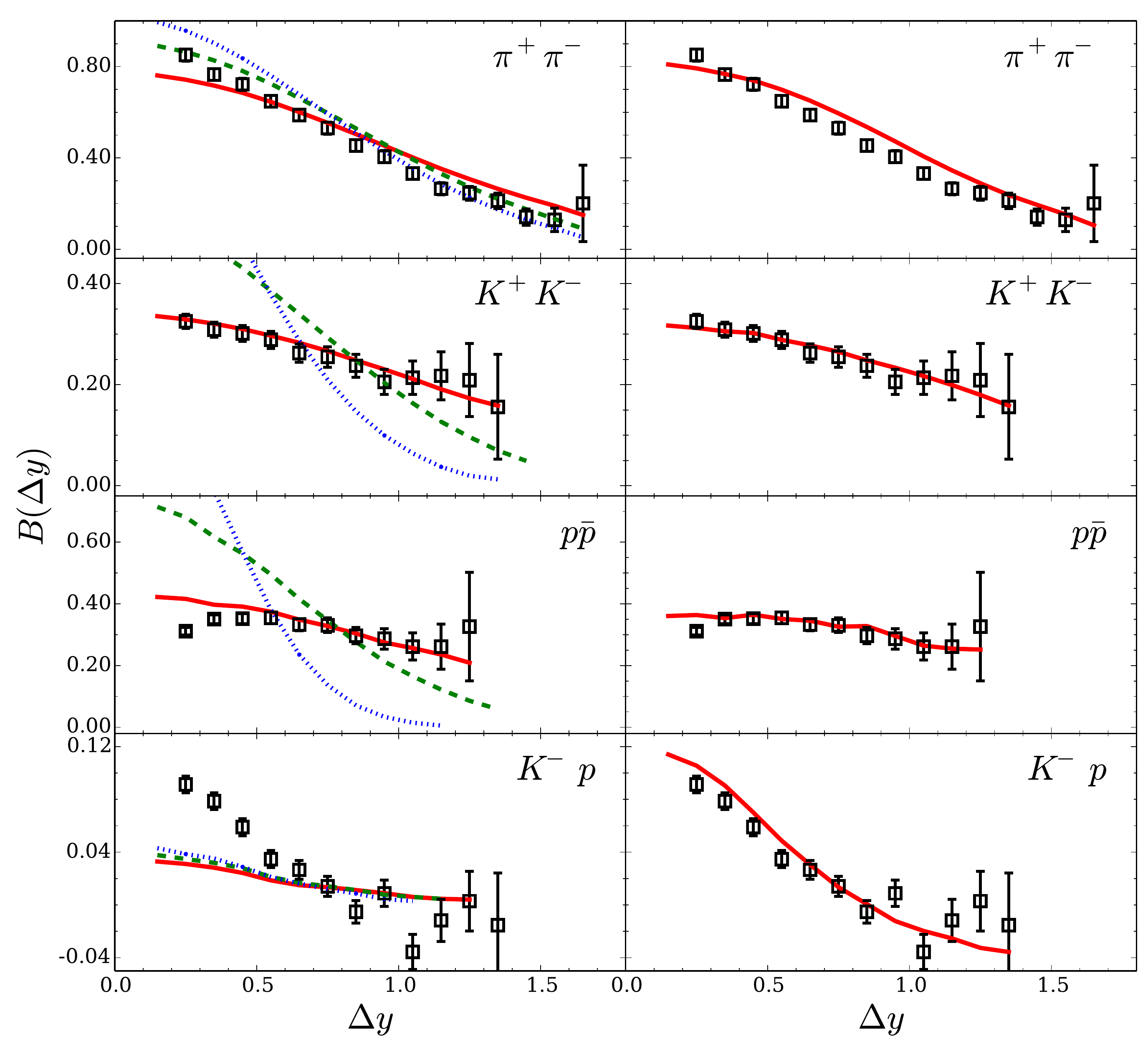}}
\caption{\label{fig:onesurgetwosurge}
Balance functions for the one-surge blast-wave model characterized by a width $\sigma_A=0.25$ (dotted blue lines), $\sigma_A=0.5$ (dashed green line) and $\sigma_A=1.0$ (solid red line) are compared to preliminary STAR results in the left-side panels. A width of $\sim 1.0$ is necessary to explain the $K^+K^-$ and $p\bar{p}$ balance functions, but this width overestimates the width of the $\pi^+\pi^-$ balance functions and fails to approach the $K^-p$ balance function. The calculations include decays that occur below $T=165$ and result in a narrow contribution that is similar but smaller than what appears in the measured $\pi^+\pi^-$ balance function.  The right-side panels show results for a two-surge model with $\sigma_A=1.0$ and $\sigma_B=0.4$. The two contributions to the $K^-p$ balance function have opposite signs which allows the resulting sum to be narrower than either the $p\bar{p}$ balance functions. The strength of the first surge is determined by matching the expected quark susceptibility to entropy ratio as extracted from lattice contributions. }
\end{figure}

The left-side panels of Fig. \ref{fig:onesurgetwosurge} compare calculations using one scale, $\sigma_A=\sigma_B$, at three values, $\sigma_A=0.25$, 0.5 and 1.0. For $\sigma_A<1.0$ the model predictions provide poor descriptions of the data. For $\sigma_A=1.0$, the $K^+K^-$ balance function is well produced and the $p\bar{p}$ balance function only differs at smaller relative rapidity. This latter failure could well be due to the lack of baryon-annihilation in the one-surge picture. The $\pi^+\pi^-$ balance function from the model is wider than the data, as it seems the model would fit better with the narrowest of the three sample balance functions. As expected, the $K^-p$ balance function cannot be explained with a single scale. As expected, the single-scale model predicts widths that fall between the widths for the $p\bar{p}$ and $K^+K^-$ balance functions. Resonances that decay into both a proton and kaon are already considered in this calculation.

\begin{minipage}{0.475\textwidth}
\begin{figure}[H]
\centerline{\includegraphics[width=0.7\textwidth]{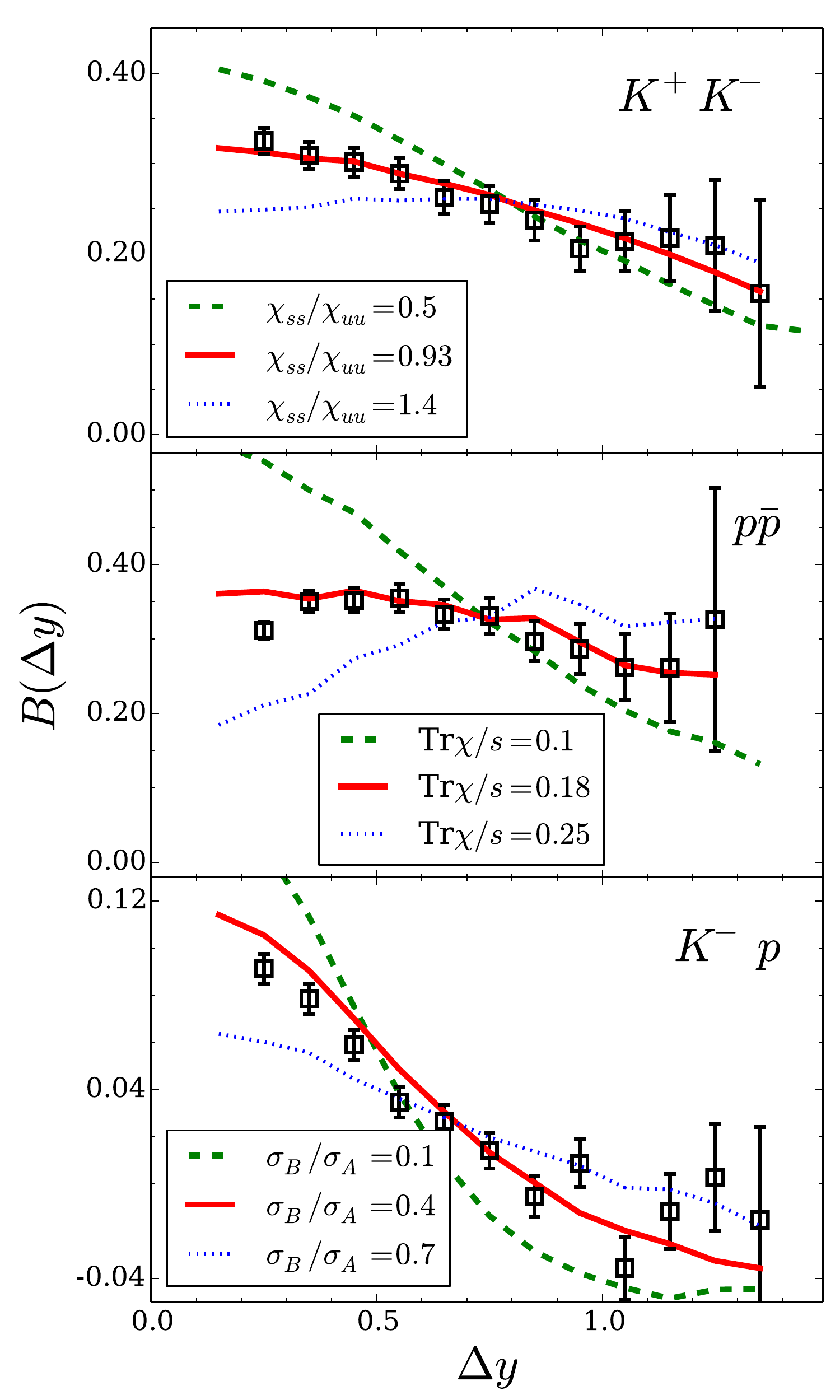}}
\caption{\label{fig:parsens}
The sensitivity of balance functions to various parameters in the two-surge blast wave model is shown above. The default calculation (solid red line) assumes the first surge achieved chemical equilibrium according to lattice calculations for $T\sim 300$ MeV. This would be a strangeness to up quark ratio, $\chi_{ss}/\chi_{uu}= 0.93$, and a net quark to entropy ratio, $(\chi_{uu}+\chi_{dd}+\chi_{ss})/s=0.18$. In the default calculation the spreads of the two surges are $\sigma_A=1.0, \sigma_B=0.4$ units of spatial rapidity. The sensitivity of the $K^+K^-$ balance function to changes in the strangeness content of the first wave is illustrated by comparing to calculations with $\chi_{ss}/\chi_{uu}=0.5$ (green dashed line) and $\chi_{ss}/\chi_{uu}=1.4$ (dotted blue line). For $p\bar{p}$, calculations were performed for different quark contents, $(\chi_{uu}+\chi_{dd}+\chi_{ss})=0.1$ (green dashed line) and $(\chi_{uu}+\chi_{dd}+\chi_{ss})/s=0.25$ (blue dotted line). $K^-p$ balance functions results are shown for varied widths, $\sigma_B=0.1$ (green dashed line) and $\sigma_B=0.7$ (blue dotted line). 
}
\end{figure}
\end{minipage}
\hspace*{0.05\textwidth}
\begin{minipage}{0.475\textwidth}
\begin{figure}[H]
\centerline{\includegraphics[width=0.7\textwidth]{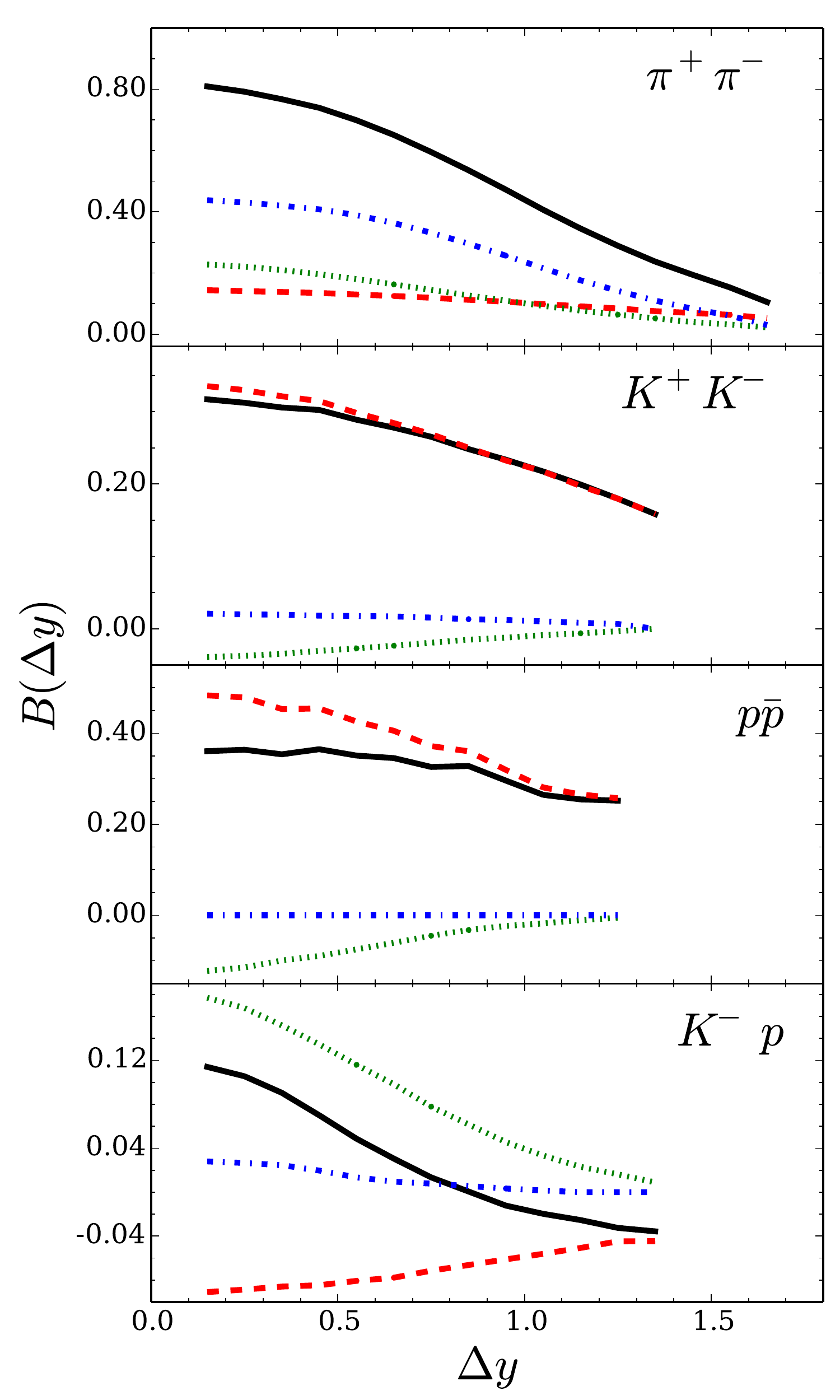}}
\caption{\label{fig:components}
Balance functions for the two-surge model with default parameters (solid black line) are divided into three components. The first surge (dashed red line) gives the widest contribution, with the width determined by $\sigma_A$ plus some thermal smearing. The width from the contribution from the second surge (dotted green line) is set by $\sigma_B$, plus thermal smearing. The third contribution is due to decays (solid blue line). The decay contribution provides nearly half the $\pi^+\pi^-$ balance function, but is much less important for the other species pairs. The $K^+K^-$ balance function is dominated by the first surge. The $p\bar{p}$ balance function has contributions from both, with the small negative contribution from the second surge resulting in a very flat balance function at small $\Delta y$. The $K^-p$ balance function has contributions of similar strength but opposite sign. This results in a narrow peak at small $\Delta y$ that is much narrower than either the $p\bar{p}$ or $K^+K^-$ balance functions. Further, at large $\Delta y$ the $K^-p$ balance turns slightly negative.
}
\end{figure}
\end{minipage}
\medskip

The two-surge model provides a far superior description of the data than a one-surge model. To illustrate the sensitivity of the predictions to specific parameters, Fig. \ref{fig:parsens} shows how some of the balance functions react to changes of the default parameterization ($\sigma_A=1, \sigma_B=0.4, (\chi_{uu}+\chi_{dd}+\chi_{ss})/s=0.18, \chi_{ss}/\chi_{uu}=0.93$). In the default parameterization the quark to entropy ratios from the first surge are set to be consistent with lattice calculations. Changing the ratio of strange to up quarks, $\chi_{ss}/\chi_{uu}$, from the first surge significantly affects the $K^+K^-$ balance function as seen in Fig. \ref{fig:parsens}. For smaller strangeness in the first surge, the second surge must then pick up the difference, which gives a larger peak at small $\Delta y$.

Adjusting the net-quark to entropy ratio, $(\chi_{uu}+\chi_{dd}+\chi_{ss})/s$, of the first surge affects all the balance functions, especially the $p\bar{p}$ balance function as shown in the middle panel of Fig. \ref{fig:parsens}. If the quark content of the first surge is below equilibrium, the second surge would have to compensate for it and the $p\bar{p}$ balance function would have a positive narrow contribution, which is not seen in the data.

Of the four balance functions studied here, the $K^-p$ balance function is the most sensitive to the separation of scales, $\sigma_B/\sigma_A$, and most strongly illustrates the need for two surges. This derives from the cancellation of the opposite-sign weights for the $A$ and $B$ contributions in Eq. (\ref{eq:Gtwoscale}). If $\sigma_A$ were to equal $\sigma_B$ the balance functions would nearly cancel as can be seen in Fig. \ref{fig:onesurgetwosurge}. As $\sigma_B/\sigma_A\rightarrow 0$, both the narrow positive peak near $\Delta y=0$, and the broad negative feature for $\Delta y\sim 1$ become more pronounced.

The $\pi^+\pi^-$ balance function is the least sensitive to changing model parameters for two reasons. First, the pions are light and their thermal motion smears any structure in relative rapidity. Second, nearly half the balancing $\pi^+\pi^-$ pairs come from resonance decays that produce both a positive and negative pion. Figure \ref{fig:components} shows the contribution to each balance function in the default two-surge calculation from the first and second surges, and from decays below the chemical freezeout temperature. The $\pi^+\pi^-$ balance function is strongly affected by decays. The decay contribution is constant in the context of this model, but could easily be sensitive to alterations of the model not considered here. In particular, this model assumes the chemical yields of various resonances do not change below chemical freezeout, $T=165$ MeV, and kinetic freezeout when the particles are emitted, $T=102$ MeV. In a realistic calculation resonances like the $\rho$ would decay and their products would rescatter before final emission. Their contribution might then move to lower values of $\Delta y$ because the relative momentum of the pions from a $\rho_0$ decay tend to be higher than those from thermal motion. Of course, the yields themselves might be different. An improved calculation would include the effects of rescattering and baryon annihilation, better account for the spectral shape of the broader mesons, and more accurately describe the acceptance and efficiency for which weak decays, e.g. the $K_s$, are captured by the detector. 

Because the strangeness susceptibility is rather flat near the transition region, and because few resonances decay to $K^+K^-$, the $K^+K^-$ balance function is dominated by the first surge. In both the preliminary STAR data and in the models, contributions from $\phi$ decays have been removed.  The $K^+K^-$ balance function is then an excellent candidate to determine $\sigma_A$. Due to baryon annihilation near and below chemical freezeout, the baryon susceptibility drops and the second-surge contribution to the $p\bar{p}$ balance function is negative. If not for thermal smearing, one would see a dip in the $p\bar{p}$ balance function. In reality, much of the annihilation occurs at the very end of the reaction and involves baryons with small relative momenta. Thus, the negative contribution from baryon annihilation should be more pronounced than the calculations shown here. The $K^-p$ balance function is especially insightful because the two contributions have similar strengths and opposite signs. For this reason, the resulting peak can be remarkably narrow, and followed by a shallow negative contribution at larger $\Delta y$. This makes it possible for modest changes in model parameters to significantly affect the size and shape of the $K^-p$ balance function. The negative dip at larger $\Delta y$ requires a large experimental acceptance. Even though STAR has an acceptance of $\pm 0.9$ units of pseudo-rapidity, the acceptance in rapidity for protons and kaons is effectively narrower due to the mapping of rapidity to pseudo-rapidity for more massive particles. Combined with the small size of the balance functions, STAR's measurements of the $K^-p$ balance function for $\Delta y>1$ is difficult.

\section{Statistical Comparison of Model to Data}
\label{sec:stats}

Figure \ref{fig:parsens} makes a case that the first surge of particle production is close to what one would expect from chemically equilibrated matter. However, the study of model responses shown in Fig. \ref{fig:parsens} is quite incomplete given that model parameters can be changed in a coordinated fashion. A more rigorous method is to simultaneously vary all model parameters while comparing to all the data. Here, we present results from not only changing the four model parameters mentioned before, but four additional parameters. The posterior likelihood is calculated in this 8-dimensional space by comparing to all four balance functions. This will provide a distribution of likely parameters that not only provides the most likely point, but provides the full likelihood in the 8-dimensional space by a Markov Chain Monte Carlo (MCMC) procedure.

\begin{table}
\begin{tabular}{|r|c|c|c|c|c|c|c|c|}\hline
parameter & $\sigma_A$ & $\sigma_B$ & $(\chi_{uu}+\chi_{dd}+\chi_{ss})/s$ & $\chi_{ss}/\chi_{uu}$ & $T_{\rm kin}$ & $u_\perp$ & $F_B$ & $\lambda_{\rm visc}$\\ \hline
Min & 0.3 & 0   & 0.05 & 0.0 & 75  & 0.5   & 0.6 & 0.7 \\
Max & 1.5 & 1.0 & 0.35 & 1.3 & 120 & 0.875 & 0.8 & 1.0\\ \hline
\end{tabular}
\caption{\label{table:parameters}
Eight parameters were simultaneously varied between the Min and Max ranges shown above. The first four parameters were two spatial rapidity widths, $\sigma_A$ and $\sigma_B$, the quark to entropy ratio of the first surge $(\chi_{uu}+\chi_{dd}+\chi_{ss})/s$, and the strangeness to up ratio of the first surge $\chi_{ss}/\chi_{uu}$. The other four parameters varied the blast wave kinematics: the kinetic freezeout temperature $T_{\rm kin}$, the transverse flow parameter $u_\perp$ defined in Eq. (\ref{eq:uperp}), the baryon reduction factor $F_B$ which accounts for baryon annihilation, and $\lambda_{\rm visc}$, which accounts for the anisotropic shape of the phase space distribution in momentum space due to viscous effects.}
\end{table}

Table \ref{table:parameters} lists the parameters varied in this analysis. The first seven parameters were described in Sec. \ref{sec:twosurgemodel} and the eighth accounts for the fact that the momentum distribution might not be anisotropic. After generating thermal particles, the momenta are scaled by a matrix
\begin{equation}
p_i=p_i+\Lambda_{ij}p_j,~~~\Lambda_{zz}=-\lambda,~\Lambda_{xx}=\Lambda_{yy}=\lambda/2.
\end{equation}
This factor reduces the thermal smearing when mapping the spatial and momentum rapidities and gives narrower balance functions. However, it does not affect the width of the decay contribution.

Even though the calculation of the balance functions requires only $\sim 10$ minutes of CPU time, hundreds of thousands of points need to be calculated during the MCMC trace. Using the techniques of model emulators in \cite{Pratt:2015zsa}, the likelihood is calculated by interpolating principal components of the data from 1536 full model runs. The full model runs were performed at points spread throughout the parameter space according to latin hyper-cube sampling. These techniques are described in \cite{Novak:2013bqa,Gomez:2012ak}.

The statistical analysis assumes Gaussian forms for the likelihood,
\begin{eqnarray}
\label{eq:lldef}
\mathcal{L}&\sim& \exp\left\{\sum_a
(y_a^{\rm(model)}-y_a^{\rm(exp)})\Sigma^{-1}_{ab}(y_b^{\rm(model)}-y_b^{\rm(exp)})/2\right\}~,\\
\nonumber
\Sigma_{ab}&=&\sigma_a^2\delta_{ab},
\end{eqnarray}
where the experimental, $y_a^{\rm(exp)}$, and model, $y_a^{\rm(model)}({\bf x})$, are compared relative to the uncertainty $\sigma_a$. The eight parameters are considered to have uniform prior distributions as listed in Table \ref{table:parameters}. The uncertainties $\sigma_a$ encapsulate not only experimental uncertainties, but any uncertainties one might expect from missing physics in the model. This can be thought of as a systematic theory error. For a schematic model such used here, assigning uncertainties is ad hoc, which means the final likelihood distribution is also suspect. However, even with that caveat, the statistical analysis helps determine what parameters best fit the data and can also assist with understanding which observables are best constraining specific parameters of interest. For this analysis, each individual point in the balance function was used, except for the first bin in relative rapidity. This bin carries both experimental difficulties due to two-track resolution and is strongly affected by femtoscopic correlations, which are ignored by the model. The uncertainty for each point was defined as 
\begin{equation}
\sigma_a=\sqrt{(0.12y_a)^2+(0.001)^2}
\end{equation}
effectively a twelve percent error. This seems rather large, but because neighboring points in a balance function provide similar information and are similarly susceptible to theoretical approximations, groups of points can be redundant. For example, if the same quantity is measured four times, a 12 percent error falls to a six percent error. So if each balance function were divided up into groups of four points, the error would translate into half this uncertainty. The MCMC method described here was repeated with several forms for the uncertainty. For smaller uncertainties the likelihoods became more compact, but the position of the maximum likelihood did not change significantly. In addition to the balance functions, the mean transverse momenta for pions, kaons and protons were treated as observables. The values were averaged from STAR and PHENIX measurements and a 6 percent uncertainty was assigned. Each balance function had 17 points, due to the binning being in units of 0.1 in relative rapidity and extending to $\Delta y<1.8$, with the first bin being neglected. The four balance functions thus provided 68 observables, adding the three mean transverse momenta result in 71 observables.

\begin{figure}
\centerline{\includegraphics[width=0.9\textwidth]{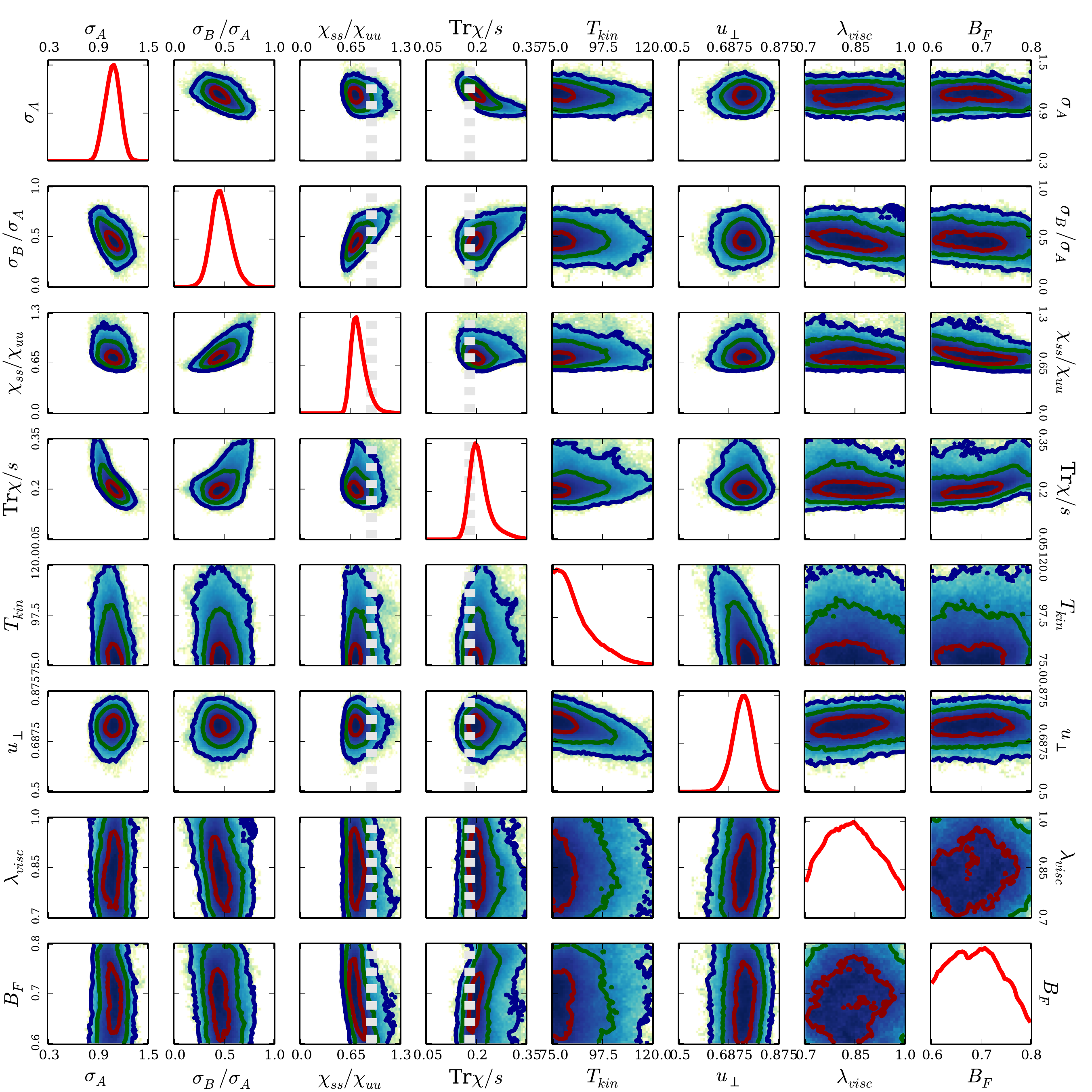}}
\caption{\label{fig:ll}
One- and two-dimensional projections of the posterior likelihood for the eight-parameter two-surge blastwave model, as computed using preliminary balance functions from STAR. Red, green and blue lines represent $1-\sigma$, $2-\sigma$ and $3-\sigma$ contours. The gray lines show lattice expectations for $\chi_{ss}/\chi_{uu}$ and $(\chi_{uu}+\chi_{dd}+\chi_{ss})/s$. The quark content of the first surge, $(\chi_{uu}+\chi_{dd}+\chi_{ss})/s$ appears about 10 percent higher than the value expected from lattice and the strangeness to up ratio is about 20\% low. 
}\end{figure}
Figure \ref{fig:ll} displays projections of the likelihood distribution as taken from the MCMC trace. The eight plots along the diagonal show one-dimensional projections, while the off-diagonal plots show two-dimensional projections. The quark content of the first surge is close to expectations from the lattice calculations shown in Fig. \ref{fig:lattice}, but might be 5\% high. Given that entropy production during the collision might be at the ten percent level, the entropy used in the denominator here might be ten percent higher than the entropy at early times. Thus, it would be not surprising if the $\chi/s$ ratios  extracted here would be 10 percent lower than the expected values for lattice calculations. This implies that the $(\chi_{uu}+\chi_{dd}+\chi_{ss})/s$ ratio may actually be $\sim 15\%$ higher than expectations and that the strangeness content, may be about 15\% lower than such expectations. This variation is at the $1-\sigma$ level in the comparison here, but because defining the uncertainty was rather arbitrary, stating a discrepancy between an extracted value and a lattice value is correspondingly arbitrary.

The widths of the two surges are strongly constrained by the analysis. The first surge's width is close to unity, which is in line with expectations. The characteristic width of the second surge appears to be just below half the broader width, $\sigma_B/\sigma_A\lesssim 0.5$. This differs from the previous analysis of \cite{Pratt:2014cza}, which found a preference for $\sigma_B/\sigma_A\lesssim 0.3$. This change is being driven by the inclusion of the $K^-p$ balance function. If neglected, there is a preference to fit the $\pi^+\pi^-$ balance function by lowering $\sigma_B$. Thus, the two balance functions are somewhat at odds with one another, which suggests that some physics is being poorly described or missing. A possible culprit would be the way in which decays are treated. If the number of decaying particles is significantly reduced between chemical and kinetic freezeout, the $\pi^+\pi^-$ balance function, which is strongly affected by decays, might narrow. The balance function with $\sigma_B/\sigma_A\sim 0.45$, would then be able to better reproduce both balance functions. 

\begin{figure}
\centerline{\includegraphics[width=0.75\textwidth]{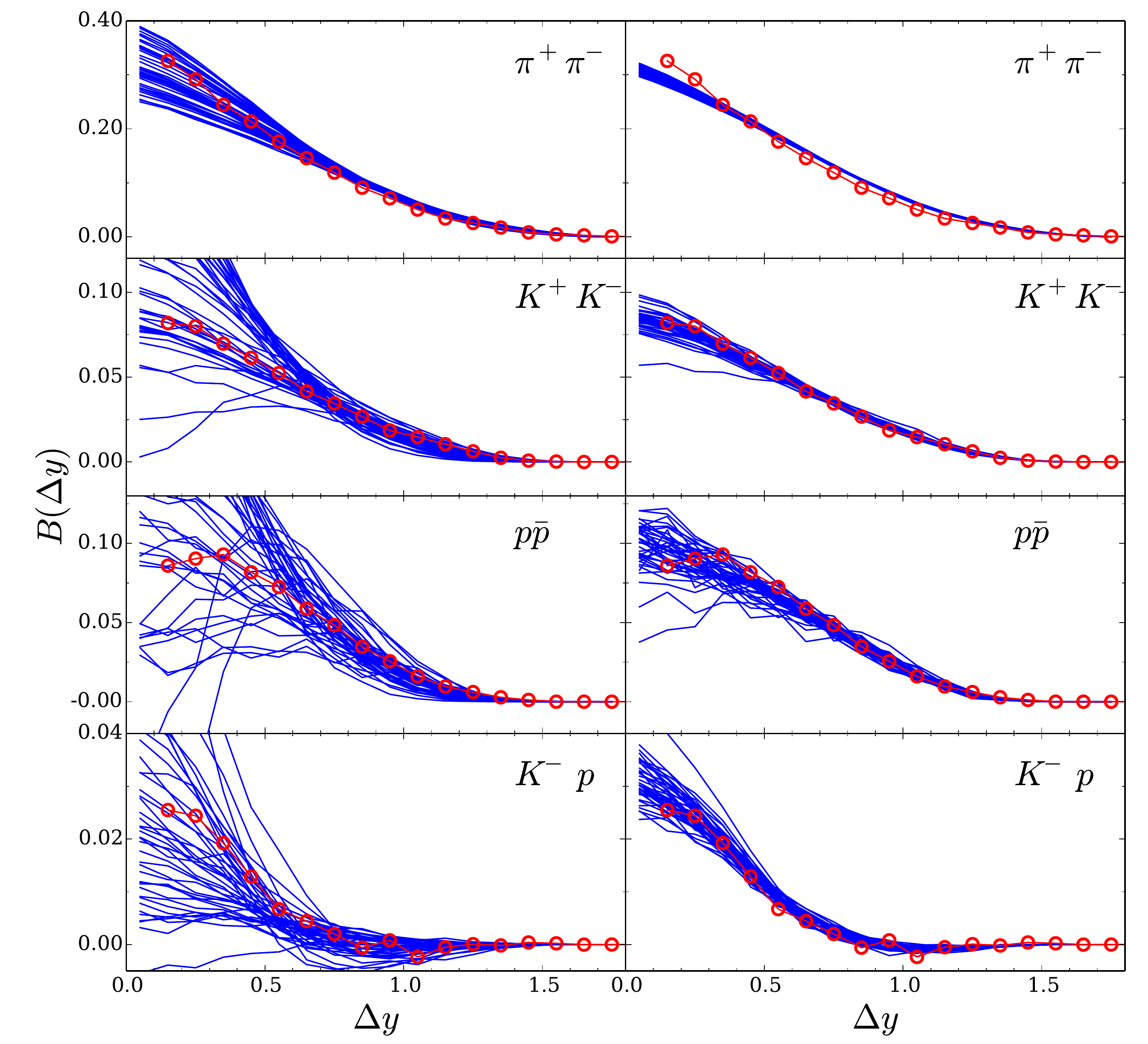}}
\caption{\label{fig:priorpost}
The left-side balance functions were calculated from 40 parameters randomly taken from the prior, i.e. they were randomly generated in the regions defined in Table \ref{table:parameters}. The parameters for the right-side balance functions were also randomly generated, but the choices were weighted with the experimental likelihood. This validates the statistical procedure and serves as a guide for how strictly the definition of uncertainties constrains the fits.}
\end{figure}

As a test of the statistical procedure we consider an assortment of balance functions calculated from random points in parameter space. In the left-side panels of Fig. \ref{fig:priorpost} the parameters are taken randomly from the prior, i.e. they are random numbers generated between the minimum and maximum values listed in Table \ref{table:parameters}. On the calculations for the right side of Fig. \ref{fig:priorpost} the parameters were taken randomly from the posterior distribution, i.e. weighted by the likelihood. These points were extracted from taking 40 points far away from one another in the MCMC trace. Whereas the balance functions on the left-side panels vary widely and often stray far from the data, the balance functions from experimentally constrained parameters on the right side come close to the data. The $\pi^+\pi^-$ balance function varies only modestly as parameters are varied throughout the prior, which means it has limited resolving power, although this could change if the $\pi^+\pi^-$ balance function turns out to be significantly sensitive to missing or poorly represented physics, e.g. reabsorption of hadronic decays into the medium. In contrast, the other three balance functions vary widely throughout the prior, as was expected from the studies presented in Fig. \ref{fig:parsens}. These measurements then have much higher potential to discriminate between different sets of parameters. The reasonably good model-to-data match validate that the statistical method was effective in identifying the most likely regions of parameter space.

\begin{figure}
\centerline{\includegraphics[width=\textwidth]{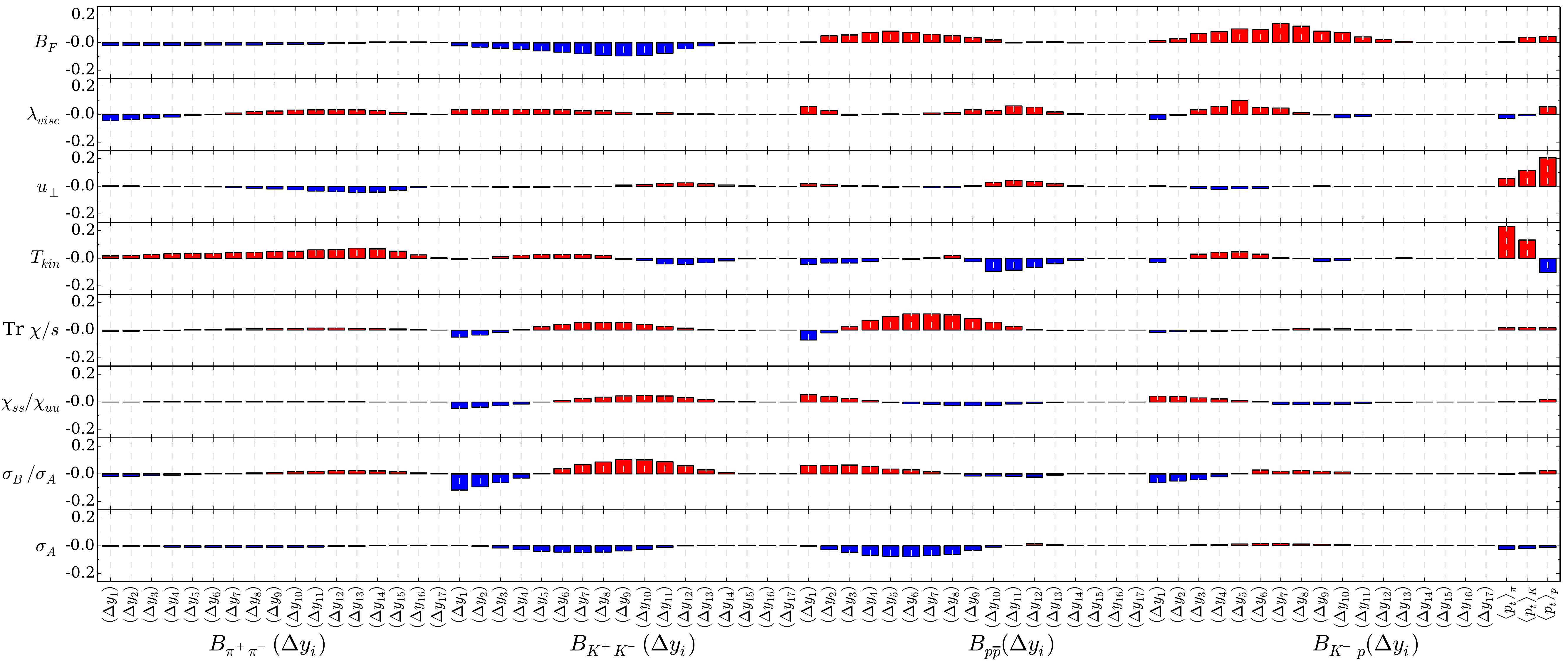}}
\caption{\label{fig:dxdy}
The ratio $\partial \langle\langle x_i\rangle\rangle/\partial y_a^{\rm(exp)}$ is shown for every combination of 71 observables and 8 parameters. This describes how an extracted value for a parameter, $\langle\langle x\rangle\rangle$, would change if a given observable were increased by one sigma while keeping the other observables fixed. Each of the four balance functions provides 17 observables, corresponding to the 18 rapidity bins with the first bin being ignored. The first 17 observables along the $x$ axis represent the $\pi^+\pi^-$ balance function, and the other three balance functions are represented by each subsequent group of 17 observables. Additionally, the three mean transverse momenta for pions, kaons and protons are included as observables. For example, one can see that the extracted value of $\chi_{ss}/\chi_{uu}$ would fall/rise if the low momentum bins of the $K^+K^-$ balance function were raised/lowered. The response is opposite for bins of higher relative rapidity. One can also see the relatively strong resolving power for the $K^+K^-$, $p\bar{p}$ and $K^-p$ balance functions. }
\end{figure}

Given the large numbers of parameters and observables, it can be difficult to understand the degree to which given measurements contribute to constraining the parameters, or the degree to which a measurement, if changed by one sigma, would lead to new parameters. In \cite{newevanpaper} methods were derived to perform sensitivity analyses from the MCMC trace. One such measure is the quantity, $\partial \langle\langle x_i\rangle\rangle/\partial y_a^{\rm(exp)}$ keeping all other $y_{b\ne a}$ fixed. This addresses the question of how the average value of a specific parameter $x_i$ in the posterior would change if a single observable $y_a$ were altered without repeating the entire MCMC with a new value of $y_a^{\rm(exp)}$. In the expression above the double brackets, $\langle\langle\cdots\rangle\rangle$, refer to an average from the posterior, and single brackets $\langle\cdots\rangle$ denote the average over the prior. Figure \ref{fig:dxdy} shows the values of this partial derivative where the parameters $x_i$ are scaled by the widths of their priors so that $\langle(x-\langle x\rangle)^2\rangle=1$, and the observables $y_a$ are scaled by their uncertainties $\sigma_a$. The resulting derivative then describes how much the parameter would change as a fraction of its prior width when the parameter $y_a$ is increased by $\sigma_a$. Figure \ref{fig:dxdy} presents a much more detailed set of information than what is shown in Fig. \ref{fig:parsens}. 

\section{Summary and Outlook}
\label{sec:summary}

Preliminary STAR measurements of charge balance functions for specific species provide strict constraints on any picture of the chemical evolution of the super-hadronic matter created in heavy ion collisions. In order to fit all four of the balance functions, the blast-wave model used here had to employ two separate surges of charge production. Balancing charges from the first surge appear to have separated by approximately one unit of spatial rapidity by the time the  matter reaches chemical freezeout. Charges from the second surge appears to have spread a bit less than half that amount. It is difficult to determine the physics driving the spread of the first surge. It may mainly be due to the separation inherent to breaking strings, or equivalently the tunneling of balancing charges in the breaking of a flux tube. Another possibility is that the spread is largely diffusive. Either way, the charges must be created early if they are to separate by such a large distance by chemical freezeout.

The strengths of the two surges can also be determined by comparing the two-surge model to data. Remarkably, the strength of the first surge is consistent with the matter from the first surge approaching chemical equilibrium. A model run based on assuming perfect equilibrium, followed by an isentropic expansion, reproduces the preliminary STAR data at the 5\% level. A more comprehensive search through eight-dimensional parameter space, including four parameters that adjust the blast-wave description, show that the best chemistry for fitting the data has susceptibilities within 20\% of those calculated on the lattice. The response of the measurements to the susceptibilities was found to be strong, and thus should provide good resolving power within the context of any model. Measurements were shown to be sensitive to both the spread and the strength of the charges.

The preliminary STAR measurements considered here are the first of their kind, and should inspire numerous additional measurements of different species, measurements as functions of relative azimuthal angle \cite{Bozek:2004dt,Schlichting:2010qia} and transverse momentum, and as a function of the three-dimensional pair momentum. As a general statement, balance functions are six-dimensional correlations that can be measured between any two species pairs. Every new combination of species and every more differential binning of the six-dimensional phase space should provide new insights and better shed light onto the story of the chemical evolution of the reaction. In addition to higher statistics, these goals may also require expanding the acceptance of current experiments.  Here, we provide a quick list of some measurements that would help illuminate an assortment issues:
\begin{itemize}
\item Measurements of unidentified particles have been made as a function of both the relative and total azimuthal angle. However, one could consider the relation between the width in azimuthal angle and relative rapidity. Balancing charges that are created early can be preferentially selected by considering the balance function at large relative rapidity, and should also have broader balance functions in relative azimuthal angle. This might help distinguish the physical cause of the separation of early charge pairs, diffusion vs. flux tube breaking.
\item The STAR Collaboration's acceptance for identified particles extends to $\pm 0.9$ units of pseudo-rapidity, but when translated into real rapidity is significantly narrower for heavier particles like kaons and protons. The ALICE detector at the LHC covers $\pm 0.8$ units of pseudo-rapidity. ATLAS and CMS cover much wider regions, but without particle identification. The long-range correlations should better illuminate the early production charges. For instance, if a measurable fraction of pairs comes from the initial state, those balancing charges might be particularly well separated. The greater acceptance would also make it easier to perform the analysis mentioned in the preceding bullet.
\item In order to better understand decays, balance functions can be binned as a function of the invariant mass or momentum. Contributions from specific decays, e.g. $K_s\rightarrow \pi\pi$, can be identified by their peak. Contribution from broader resonances such as $\rho_0\rightarrow\pi^+\pi^-$ may be more difficult to see, but nonetheless the comparison should be able to identify whether the populations of unstable resonances are poorly represented by this, or any other, model. 
\item Measurements can be performed as a function of beam energy and centrality \cite{Adams:2003kg} and for different choices of colliding nuclei \cite{Aggarwal:2010ya}. The narrowing of the balance function for unidentified particles may be related to the existence of two surges \cite{Bass:2000az}, but that conclusion needs to be investigated in detail alongside the analyses listed above. 
\end{itemize}

If these measurements are to lead to truly rigorous conclusions, the models must improve beyond the schematic blast wave model considered here. In order to justify Gaussian forms for the two surges, each surge must be created during a short time. More realistically, charges are created continuously according to the rate at which $\chi_{ab}/s$ is changing. The evolution of the charges and the creation of new charge pairs after chemical freezeout is an area that was poorly modeled here. By the time of kinetic freezeout many of the resonances will already have decayed, with their products being reabsorbed by the medium. Baryon annihilation, which was implemented here by reducing the baryon yield at chemical freezeout, in fact occurs mostly later. The corresponding dip should then be more concentrated at low relative momentum, a feature that is indeed seen in the data. These improvements cannot be naturally added to a schematic blast wave model, but they can be accounted for in microscopic simulations. Such calculations are entirely tractable and are being performed currently by this group. Once the models can encapsulate all the important physics, statistical analyses like those presented here can lead to much more convincing and rigorous conclusions.


\begin{acknowledgments}
The authors thank Hui Wang for providing STAR data, and Gary Westfall for providing routines to simulate STAR's acceptance. This work was supported by the National Science Foundation's Cyber-Enabled Discovery and Innovation Program through grant NSF-0941373 and by the Department of Energy Office of Science through grant number DE-FG02-03ER41259. The work of C. R. is supported by the National Science Foundation through grant number NSF PHY-1513864.
\end{acknowledgments}

\end{document}